# TOWARDS TO AN AGENT-ORIENTED MODELING AND EVALUATING APPROACH FOR VEHICULAR SYSTEMS SECURITY

Mohamed GAROUI[1], Belhassen MAZIGH[2], Béchir El AYEB[3], Abderrafiaa KOUKAM[4]

[1]PRINCE Research Unit, National School of Computer Science, Manouba, Tunisia
[2]Department of Computer Science, Faculty of Sciences Monastir, Tunisia
[3]PRINCE Research Unit, Faculty of Sciences Monastir, Tunisia
[4]IRTES-SET, EA 7274, UTBM, F-90010 Belfort cedex, France

## ABSTRACT

*Agent technology is a software paradigm that permits to implement large and complex distributed applications. In order to assist the development of multi-agent systems, agent-oriented methodologies (AOM) have been created in the last years to support modeling more and more complex applications in many different domains. By defining in a non-ambiguous way concepts used in a specific domain, Meta modeling may represent a step towards such interoperability. In the Transport domain, this paper propose an agent-oriented meta-model that provides rigorous concepts for conducting transportation system problem modeling. The aim is to allow analysts to produce a transportation system model that precisely captures the knowledge of an organization so that an agent-oriented requirements specification of the system-to-be and its operational corporate environment can be derived from it. To this end, we extend and adapt an existing meta-model, Extended Gaia, to build a meta-model and an adequate model for transportation problems. Our new agent-oriented meta-model aims to allow the analyst to model and specify any transportation system as a multi-agent system. Based on the proposed meta-model, we proposes an approach for modeling and evaluating the Transportation System based on Stochastic Activity Network (SAN) components. The proposed process is based on seven steps from "Recognition" phase to "Quantitative Analysis" phase. These analyzes are based on the Dependability models which are built using the formalism Stochastic Activity Network. A real case study of Urban Public Transportation System has been conducted to show the benefits of the approach.*

## 1. INTRODUCTION

The purpose of the Agent-Oriented Software Engineering is the creation of a path towards integration and interoperability of methodological approaches for multi-agent systems (MAS) development. This involves the definition of a common framework for MAS specification, which includes the identification of a minimum set of concepts and methods that can be agreed in the different approaches. The tool for defining this framework is meta-modelling. The principle of meta-modelling has been already used in other fields of software engineering, for instance, in the specification of UML [1] by OMG, to describe the elements of the language, their constraints and relationships.

In platooning systems Research such as in [2, 3], each vehicle determines its own position and orientation only from its perceptions of the surrounded environment. In this context, the reactive multi-agent paradigm is well adapted to specify and analyze this system. The interest of those approaches results from their adaptability, simplicity and robustness. In this case, platoon configuration can be considered as the result of the self-organization of a reactive multi-agent system (RMAS). A platoon multi-agent system can then be defined as a set of agents, each one corresponding to a vehicle. Two agent roles can be distinguished: leader and follower agents.

Our problem here is when we model vehicular system, we need an agents-oriented meta-model that gives us a set of basic concepts. These concepts are necessary to model the entire of transport problem in different *environment* (Urban, Agricultural, and Military) and with various *navigation policies* and its *behaviour* describing the *system scenarios*. In addition, as soon as we obtain the system model, it will be easy to implement our multi-agent system by using agent oriented programming.

In this paper, our contribution is to provide an agent-oriented meta-model adequate to transportation problem which allowed us to model the vehicular platoon and any others transportation problem in their navigation environment. Our proposed meta-model has been built by *adopting* and *extending* the existing *Extended Gaia meta-model* [4] and thus we define two levels of models inspiring from PASSI meta-model [5]. This seems to us coherent with the most accepted definition of meta-model: a meta-model is a "model of a model", and it provides an explicit representation of the constructs and relationships needed to build specific models within a domain of interest. This proposition arises by remarking that in the field of transport doesn't occur any Agent oriented meta-model to clearly specify and analyse any transport system in the form of multi-agent systems. By relying on our proposed meta-model, we propose a process for modelling and evaluating the system dependability (we are interested to the system security which is related to system availability and reliability) using SAN [5, 6] formalism. The aim of this contribution was to evaluate the impact of failures and disturbances of system upon the overall system security. In our work, we define the notion of the *system security* by: "*secure system must monitor and evaluate the safe-security system itself and all the internal and external entities related to the system in different traffic conditions and in diverse environment (Urban, Agricultural and Military). I.e. the safety of our system is defined by that the system be in made available in all times on the traffic lane and does not cause any accidents or catastrophic consequences for the environment*".

We choose to use the Extended Gaia meta-model as it is well adapted to organizational structures such as *teams*, *congregations* and *coalitions* which are used in clustering and collaborative missions of the platoon entities. Furthermore, the proposed approach must take into account, in their meta-model, the concept of environment and different social structures associated with different application areas (Urban, Agricultural, and Military) as indicated in Table 1. Extended Gaia specifies the notion of the environment by Environment concept. The abstraction of the environment specifies the set of entities and resources of a multi-agent system can interact with, limiting interactions using the authorized shares.

Table 1. Social structure according to the application areas.

| Application Area | Suitable Social Structure |
|---|---|
| Urban | Congregations, Coalition |
| Agricole | Congregations, Teams, Coalition |
| Military | Teams, Congregations, Coalition |

The *Extended Gaia* meta-model adds some organizational based concepts. The organization itself is represented with an entity, which models a specific structure (or topology). The organizational rules are considered responsibilities of the organization. They include safety rules (time-independent global invariants that the organization must respect) and a liveness rules (that define how the dynamics of the organization should evolve over time).

This paper is structured as follow: in Section 2, we present same related works about the existing agent-oriented meta-model used for modelling and specify multi-agent systems. Section 3 presents our Proposed Agent-oriented Meta-model for transportation systems. Then, Section 4 depicts the proposed methodology for evaluating Transportation System. Sections 5 illustrate our modelling approach with an application of urban public transportation systems. Finally, Section 6 concludes by giving a list of possible future works

## 2. RELATED WORKS

Many agent-oriented meta-model have been proposed for modelling of multi-agent system. The first version of the Gaia methodology, which modelled agents from the object-oriented point of view, was revisited 3 years later by the same authors in order to represent a MAS as an organized society of individuals [6, 7].

Agents play social roles (or responsibilities) and interact with others according to protocols determined by their roles. With that approach, the overall system behaviour is understood in terms of both micro- and macro-levels. The former explains how agents act according to their roles, and the latter explains the pattern of behaviour of those agents. These constraints are labelled organization rules and organization structures respectively.

A central element of the meta-model of Gaia is the agent entity, which can play one or more roles. A role is a specific behaviour to be played by an agent (or kind of agents), defined in terms of permissions, responsibilities, activities and interactions with other roles. When playing a role, an agent updates its behaviour in terms of services that can be activated according to some specific pre- and post-conditions. In addition, a role is decomposed in several protocols when agents need to communicate some data. The environment abstraction specifies all the entities and resources a multi-agent system may interact with, restricting the interactions by means of the permitted actions.

The Extended Gaia meta-model adds some organizational based concepts. The organization itself is represented with an entity, which models a specific structure (or topology). The organizational rules are considered responsibilities of the organization. They include safety rules (time-independent global invariants that the organization must respect) and liveness rules (that define how the dynamics of the organization should evolve over time). Given the aggregation association defined from agents with respect to organizations and from organizations with respect to organization structures, Gaia permits to design a hierarchical non-overlapping structure of agents with a limited depth. From the organizational point of view, agents form teams as they belong to a unique organization, they can explicitly communicate with other agents within the same organization by means of collaborations, and organizations can communicate between them by means of interactions. If inter-organization communication is omitted, coalitions and congregations may also be modelled.

**PASSI** (Process for Agent Societies Specification and Implementation) [8] is an iterative-incremental process for designing multi-agent systems starting from functional requirements that adopts largely diffused standards like UML (as the modelling language, although extended to fit the needs of agents design) and FIPA (as the agent platform). PASSI covers all the phases from requirements analysis to coding and testing with a specific attention for the automation of as many activities as possible with the support of PTK (PASSI Toolkit), a specifically conceived design tool.

The PASSI MAS meta-model [8] is organized in three different domains: the Problem Domain (where requirements are captured), the Agency Domain that represents the transition from problem-related concepts to the corresponding agent solution (that is a logical abstraction), and the Solution Domain (where the implemented system will be deployed).

The Problem Domain deals with the user's problem in terms of scenarios, requirements, ontology and resources; scenarios describe a sequence of interactions among actors and the system. Requirements are represented with conventional use case diagrams. The system operating environment is depicted in terms of concepts (categories of the domain), actions (performed in the domain and effecting the status of concepts) and predicates (asserting something about a portion of the domain elements), the environment also includes resources that can be accessed by agents.

The Agency Domain includes the agent that is the real centre of this part of the model; each PASSI agent is responsible for accomplishing some functionalities descending from the requirements of the Problem Domain. Each agent during its life can play some roles; these are portions of the agent social behaviour characterized by some specificity such as a goal, or providing a functionality/service and in so doing it can also access some resources. The Service component represents the service provided by a role in terms of a set of functionalities (including pre- and post-conditions as well as many other details mostly coming from the OWL-S specifications), and can be required by other agents to reach their goals. Agents could use portions of behaviour (called tasks) or communications to actuate the roles aims.

**ASPECS** (Agent-oriented Software Process for Engineering Complex Systems) provides a holonic perspective to design MAS [9]. Considering that complex systems typically exhibit a hierarchical configuration, on the contrary to other methodologies, it uses holons instead of atomic entities. Holons, which are agents recursively composed by other agents, permit to design systems with different granularities until the requested tasks are manageable by individual entities.

Being one of the most recent methodologies, it takes the experience gained from previous approaches (such as PASSI and RIO [10] as the base to define the meta-model and the methodology.

The goal of the proposed meta-model is to gather the advantages of organizational approaches as well as those of the holonic vision in the modelling of complex systems. A three layer meta-model, with each level referring to a different aspect of the agent model, is proposed: The Problem domain covers the organizational description of the problem. An organization is composed by roles which interact within scenarios while executing role plans. Roles achieve organizational goals by means of their capacities (i.e., what a behaviour is able to do). The organizational context is defined by means of ontology. This meta-model layer is used mainly during the analysis and design phases. The Agency domain defines agent-related concepts and details the holonic structure as a result of the refinement of the elements defined in the Problem domain. Each holon is an autonomous entity with collective goals and may be composed by other holons. Holonic groups define how members of the holon are organized and how they interact in order to achieve collective goals. At the finest granularity level, holons are composed by groups and their roles are played by agents, which achieve individual goals. A rich communication between agent roles (which are instances of organizational roles) is also supported, specifying communicative acts, knowledge exchange formalized by means of the organizational ontology, and protocols specifying sequences of messages.

## 3. OUR PROPOSED META-MODEL: PLATOONING META-MODEL

UML is based on the four-level meta-modelling architecture. Each successive level is labelled from $M_3$ to $M_0$ and are usually named meta-meta-model, meta-model, class diagram, and object diagram respectively. A diagram at the $M_i$-level is an instance of a diagram at the $M_{i+1}$-level. Therefore, an object diagram (an $M_0$-level diagram) is an instance of some class diagram (an $M_1$-level diagram), and this class diagram is an instance of a meta-model (an $M_2$-level diagram). The $M_3$-level diagram is used to define the structure of a meta-model, and the Meta Object Facility (MOF) belongs to this level. The UML meta-model belongs to the $M_2$-level.

After studying the Extended Gaia meta-model, we observe how this explicit and useful models of the social aspect of agents. Although it was not designed for open systems, and provides little support for scalability, simplicity allows improvements to facilitate with a relative. it models both the macro and micro aspects of the multi-agent system. Gaia believes that a system can be regarded as a company or an organization of agents.

In this section, we try to solve our contributions mentioned from the start. It manifests itself to extend and adapt an existing meta-model to build a meta-model and an adequate model for transport problems.

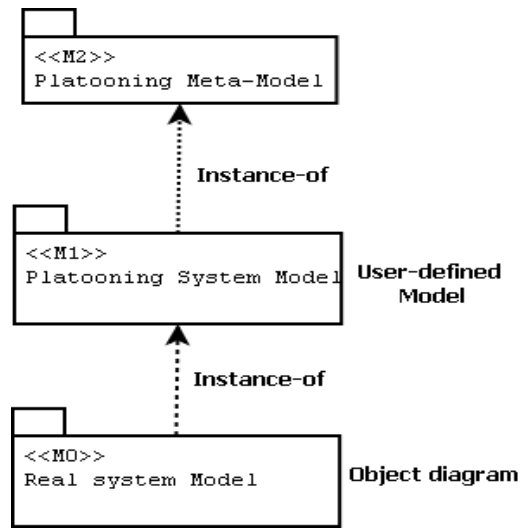

Figure 1. Model instantiation checking

In Fig. 2, the classes present in black colour are the base classes of Extended Gaia meta-model. For against, the blue classes are the classes added to the existing meta-model to be adapted to platooning applications and then help us to implement our own methodology for modelling and dependability analysis. Table 2 presents the definition of the added new concepts.

Table 2.  Definition of the new added concepts.

| Concept | Definition |
|---------|------------|
| Functional Requirement | A function that the software has to exhibit or the behaviour of the system in terms of interactions perceived by the use |
| Non-Functional Requirement | A constraint on the solution. Non-functional requirements are sometimes known as constraints or quality requirements |
| AgentModel | Abstract description of a formal model which gives an abstract view about ***the agent behaviour***. |
| OrganizationModel | Abstract description of a formal model which gives an abstract view about ***the organization behaviour***. |

The concept Functional Requirement is a function that the software has to exhibit or the behaviour of the system in terms of interactions perceived by the use. This concept allowed us to identify our system requirements. The Non-Functional Requirement concept provides a constraint on the solution. Non-functional requirements are sometimes known as constraints or quality requirements. *AgentModel* concept gives an abstract view about **the Agent behaviour**. *OrganizationModel* gives an abstract view about **the organization behaviour**.  Behaviour is described by formal state-based models [11].

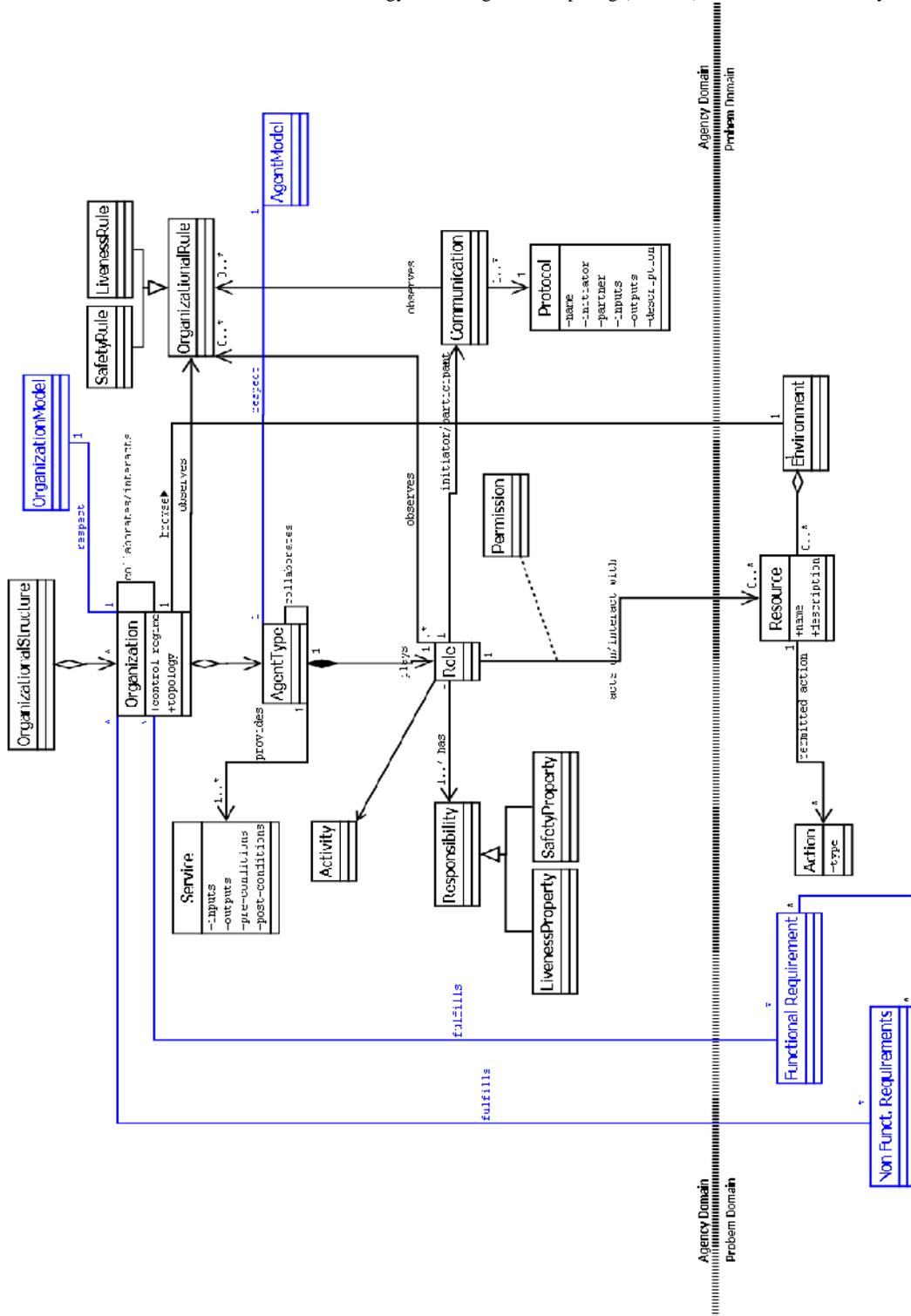

Figure 2. Our Meta-model: Platooning Meta-model.

By inspiring from PASSI [8] and ASPECS meta-models [9], we tried to organize our meta-model in two areas: Problem Domain and Agent Domain. Problem Domain involves elements (Fig. 2) are used to capture the requirements problem and perform initial analysis. Agent Domain includes elements (Fig. 2) are used to define an agent-oriented solution to the problem described in the previous step.

After this, we pass to $M_1$-level describes the Platooning System Model (see Fig. 3) which constitutes of instance of the concepts of $M_2$-level model. This model includes all the basic concepts and necessary for us to model any type of application to platooning with their bodies, interaction, environment, their geometric configuration and formal models associated with each component platoon. The table below provides the concepts related to platooning System Model and their relationship with the concepts of the $M_2$-level model.

Table 3.  The related concepts to Platooning System Model.

| Concept | Instance of |
|---|---|
| Platoon | Organization |
| Structure | OrganizationalStructure |
| Geo_Configuration | OrganizationalRule |
| Navigation_Policy | OrganizationalRule |
| Interaction | Communication |
| Entity | AgentType |
| Leader | AgentType |
| Follower | AgentType |
| Parameters | OrganizationalRule |
| Entity_Parameters | -- |
| System_Parameters | -- |
| Model | -- |
| Entity_Model | AgentModel |
| Platoon_Model | OrganizationModel |
| Area | Environment |

The table below gives an idea about the basic concepts of *Platooning System Model* (Fig. 3) which is instances of our meta-model that shown in the Fig. 2. The *Platoon* concept represents the main element in our model which is an instance of meta-concept *Organization*. Any Platoon is modelled as a set the *Entity*. There are two kind of entity: Leader and Follower which are modelled by the two concepts *Leader* and *Follower*. The tow concepts *Entity_Model* and *Platoon_Model* are used to describe the behaviour of entity and platoon in the environment. The concept *Area* model the environment notion. In our transportation problem, there are three types: Urban, Agricultural, and Military. The concepts *Parameters*, *Entity_Parameters* and *System_Parameters* provided a general idea about the parameters of the entities and of the system. These parameters are necessary and useful for Dependability Evaluation in our future work.

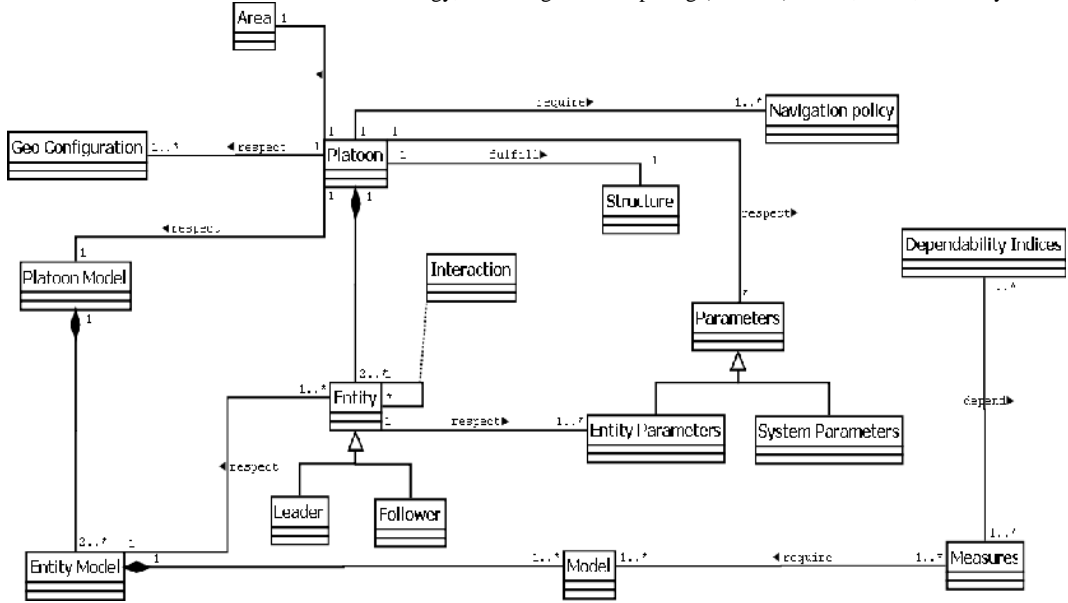

Figure 3. Platooning System Model

## 4. PROPOSED METHODOLOGY OVERVIEW

The nation's transportation system quickly, safely, and securely moves people and goods through the country and overseas. The transport systems are used in various fields: urban, agricultural and military and are also useful for performing activities in a well specific field for example the passengers transfer in urban areas. For a model-based analysis and Dependability evaluation of transportation system in a dynamic environment, we propose the methodology described on Fig. 4 that sketches the main process steps.

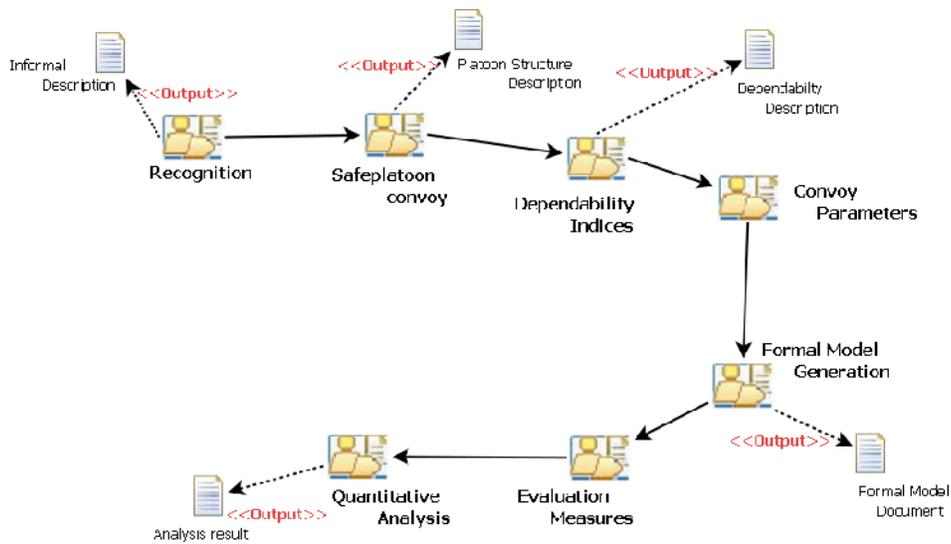

Figure 4. Platooning System Model

**Step 1**-*Recognition*: The first activity in the analysis is to conduct a reconnaissance of any convoy. The objective here is to determine *system structure*, *navigation policy* in the its *environment* and the *adopted strategies* by the convoy studied. Furthermore, different scenarios can be defined, which can then serve as a starting point for finding a good configuration that is compatible with the reality of convoys.

**Step 2**-*Safeplatoon convoy*: The result of the previous step is a convoy description. In this activity, we have the results of the previous step in order to specify the number of each entity involved in the convoy and the clustering strategy and communication between vehicle convoy. The following entities are frequently observed in various types of convoys Safeplatoon:

- Vehicle **Header**: For each type of convoy vehicle plays the leading role and can be driven by an operator or be independent. For column configurations and level the leader is usually the first vehicle in the direction of movement of the convoy. If vehicles are numbered according to their position, the index of the leader is i = 1.
- Vehicle **Follower**: These are vehicles connected by a virtual link to a header vehicle.
-

**Step 3**-*Dependability Indices*: In this phase, we set the dependability indicators of our system. There are several dependability indexes to analyze and evaluate the studied system.

**Step 4**-*Convoy Parameters*: After defining dependability indicators, it is necessary to set the parameters of the convoy that will be examined and used to generate a further analyzable model. There are two types of parameters: the entities parameters (*Entity_parameters*) and system parameters (*System_parameters*). Within the parameters of the entities, we fix the maximum speed, the acceleration/deceleration and the mass of the entity. And the parameters of the system must determine the number of entities, geometric configuration of the convoy, the lateral gap (between the entities) and the longitudinal gap.

**Step 5**-*Formal Model Generation*: In this phase, we are interested in the formal model describing the behavior of leading in convoy in a specific dynamic environment. This model is constructed using a formal language [12].

**Step 6**-*Evaluation Measures*: If the formal model of the convoy is valid, chosen dependability indices must be converted to measures which are calculated from the model. The main measures are: MTTR and MTTF.

**Step 7**-*Quantitative Analysis*: The refined model should now be analyzed by considering the chosen evaluation method (simulation method or analytic method). This assessment must calculate the metric results. Once the results obtained, it is necessary to analyze them.

System formal model is constructed using a formal language. It describes the system behavior with/without externs and internal events that disrupt the system in the dynamic environment. The disrupted system avoids catastrophic consequences on the user(s) and the environment.

Different scenarios can then be evaluated. The scenarios are constructed by varying the model parameters (e.g. the number of available vehicles, geometric pattern, etc.) or by changing the structure of the convoy. If it is not necessary to make changes to the convoy, new parameters can be applied directly to the refined formal model. The analyses of the results of different scenarios help to find out what are the best.

## 5. A CASE STUDY: URBAN PUBLIC TRANSPORTATION SYSTEM

According to the Agent-Oriented Meta-model, we try to specify a transport applications in urban environment. The convoy adopts a *line* configuration with Longitudinal gap (Inter-distance) between vehicle 0 meter and 2 meters in lateral gap. For these scenarios, the convoy will have a fixed number of vehicles between two and three and will move on a track with a radius of curvature ranging from 15 m to infinity. The train moves at a maximum speed of 15 km/h with an acceleration of 1 m/s$^2$ and a deceleration of -3 m/s$^2$ on a maximum distance of 1000 meters. From these parameters, two scenarios are proposed. The first is to evolve a convoy of vehicles with fixed Line configuration (see Fig. 5a). During the movement, the convoy can change its geometric configuration from Line configuration to Echelon configuration (see Fig. 5).

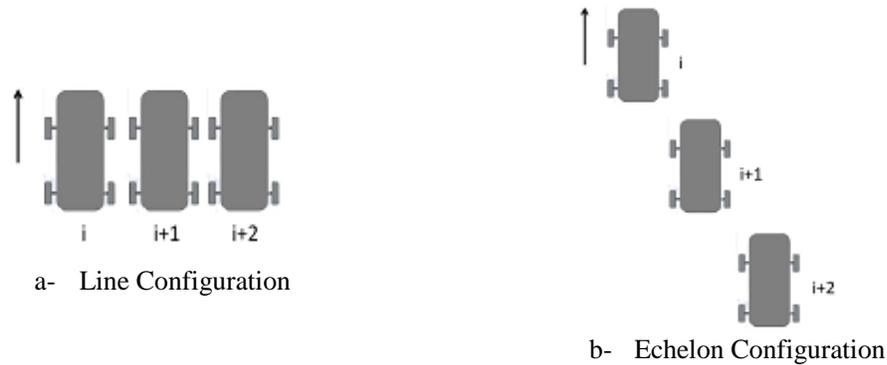

a- Line Configuration

b- Echelon Configuration

Figure 5. Platoon configurations

In Fig. 6, we present the object diagram which is an instantiation of the *Platooning System Model* (Fig. 3). The object diagram in the data modelling language UML used to represent instances of classes, that is to say objects. As the class diagram, it expresses the relationship between objects, but also the state of objects, thereby expressing execution contexts. In this sense, this pattern is less general than the class diagram. Object diagrams are used to show the state of object instances before and after the interaction, i.e. it is a photograph at a specific time and attributes existing object. It is used in the exploratory phase.

The object diagram of our study is a set of objects that have the attributes that characterize the system. The object *Convoy_Urban* is an instance of the concept *Platoon* which is it's an instance of *Organization* concept of the metamodel. The convoy adapt a line configuration and Line to Level navigation policy therefore we find an instance of *Geo_Configuration* named *Line_Configuration* and an instance of *Navigation policy* named *Line_to_Level*.

Our transportation system is constitutes of three intelligent vehicles: one Leader and two follower, thus the object diagram contains two items: *V_Leader* instance of the concept *Leader* with cardinality equal to 1 and *V_Follower* instance of *Follower* concept with cardinality equal to 2.

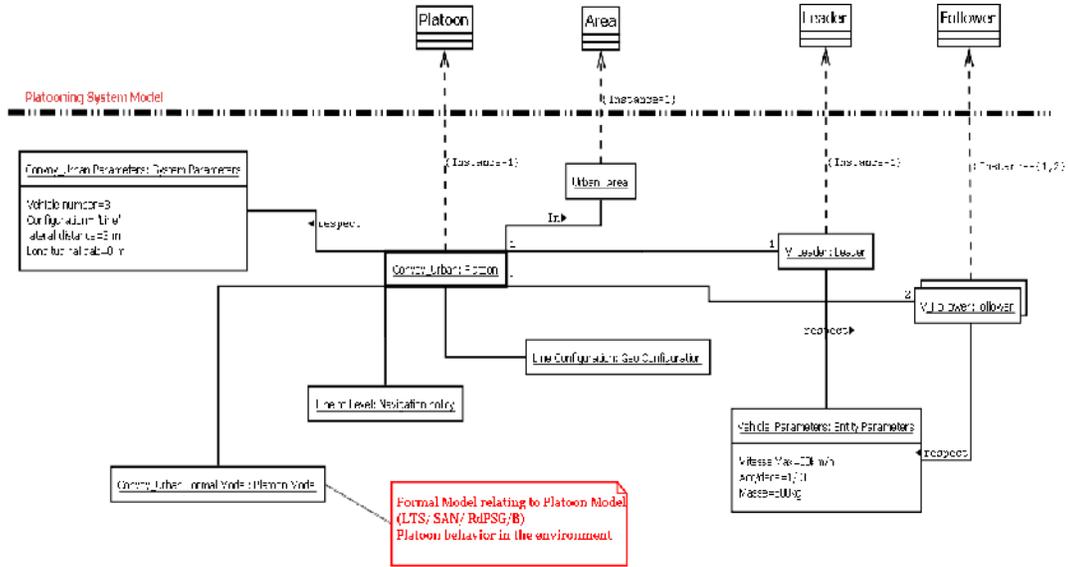

Figure 6.  Object diagram relating to urban public transportation systems

Our system has some parameters regrouped in the two tables 4 and 5. These parameters are divided into two kinds: *Vehicle_Parameters* and *Convoy_Urban_Parameters* which respectively represent convoy entities parameters and the parameters of the overall system. They are used for dependability evaluation in our future works. Transport system behaviour is modelled by *Convoy_Urban_Formal_Model* object. The behaviour is described by state-based models which are used in system dependability evaluation. This model and parameters are used in our future work to the dependability evaluation.

Table 4.  Vehicles Parameters.

| Parameters | Values |
|---|---|
| Max speed | 50 km/h |
| Acceleration/ deceleration | 1 m/s$^2$/-3 m/s$^2$ |
| Weight | 500 kg |

Table 5.  Convoy Parameters.

| Parameters | Values |
|---|---|
| Vehicle Number | 3 |
| Configuration | "Line" |
| lateral gap | 3 m |
| longitudinal gap | 0 |

To validate our proposed evaluation process, we try to specify this example of transport application in urban environment.

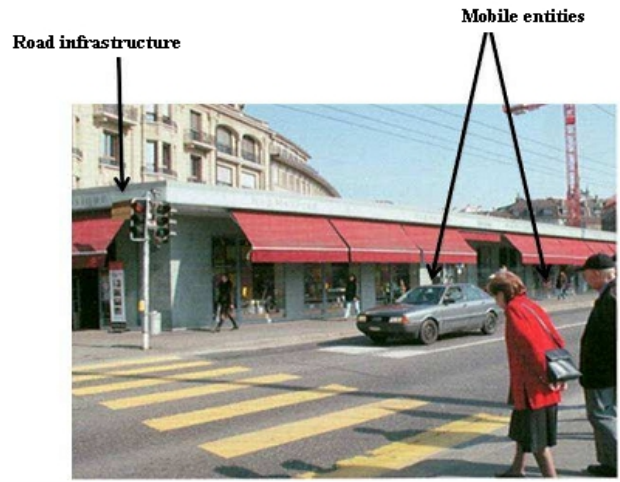

Figure 7. Dynamic urban environment

The urban environment (see Fig. 7) can also be seen as a dynamic and instable. A set of external events may intervene on the traffic lane. These events can come from mobile entities in the environment and the existing road infrastructure: Mobile entities: the traffic lane can be disturbed by mobile entities. An entity can be a pedestrian, bicycle, motorcycle or car. These entities may be on the way and in influence the traffic on it. A disturbance of this type can cause a stoppage or reduction in speed of the convoy. Road infrastructure: the traffic lane is equipped with a dynamic road infrastructure. It is composed of a set of elements that can cause stoppages or slowdowns vehicles on the road. Lights, intersections can at any time force stop the convoy. The system is considered operational if and only if there are no external events in the traffic lane. In what follows, we describe the steps of our methodology applied onto our case study. In what follows, we describe the steps of our methodology applied onto our case study.

***Step 1****-Recognition*
The first activity in the analysis of the convoy is to conduct a reconnaissance of any convoy. Our convoy is a public transportation system in an urban environment. It adopts a line configuration and follows a line to level navigation policy (*Line_to_Level*).

***Step 2****-Safeplatoon convoy*
In this phase, our goal is to specify the number of each entity involved in the convoy and the clustering strategy and telecommunication links between vehicle convoy. Our convoy will have a fixed number of vehicles equal to 3. Vehicles are of two types: a header (*Leader*) and two followers (*Follower*).

***Step 3****-Dependability Indices*
In this phase, we set the dependability indicators related to the system. There are several dependability indicators to analyze and evaluate any systems. Mention for example the MTTR (Mean Time To Repear) and the MTTF (Mean Time To Failure)

***Step 4-****Convoy Parameters*

In this phase, it is necessary to set the entities and system parameters. These parameters will be discussed over and are used to generate an analyzable model for the system. Our convoy adopts a longitudinal gap of 0 meters and lateral gap of 2 meters and move on a track with a radius of curvature ranging from 15 m to infinity. The convoy moves at a maximum speed of 15 km/h with an acceleration of $1ms^2$ and a deceleration of $3ms^2$ on a maximum distance of 1000 meters. The two tables 6 and 7 represent the parameters of entities and system settings.

Table 6.  Vehicles Parameters
.

| Parameters | Values |
|---|---|
| Max speed | 50 km/h |
| Acceleration/ deceleration | $1 \text{ m/s}^2$/-3 m/s$^2$ |
| Weight | 500 kg |

Table 7.  Convoy Parameters.

| Parameters | Values |
|---|---|
| Vehicle Number | 3 |
| Configuration | "Line" |
| Inter-distance | 3 m |
| longitudinal gap | 0 |

***Step 5-****Formal Model Generation*

During this phase, we are interested in the formal model describing the leading in convoy behavior in a fixed environment. In our case, we chose to use the SAN formalism to build the dependability model associated to the duct in a convoy in the urban environment. SAN models describe the nominal behavior of convoy, i.e. in situations faultless and exceptional conditions, namely the failure modes. Fig. 8 shows the behavioral model of our convoy. This model is motto on three modules: *nominal behavior*, *faulty behavior in the presence of a pedestrian or infrastructure* and *faulty behavior in the presence of a mechanical failure in the convoy*.

The nominal behavior represents the non-defaulting behavior of the system by against the two others modules respectively model faulty behavior in the presence of a pedestrian or infrastructure in the traffic lane and the faulty behavior in the presence of a mechanical failure in the convoy.

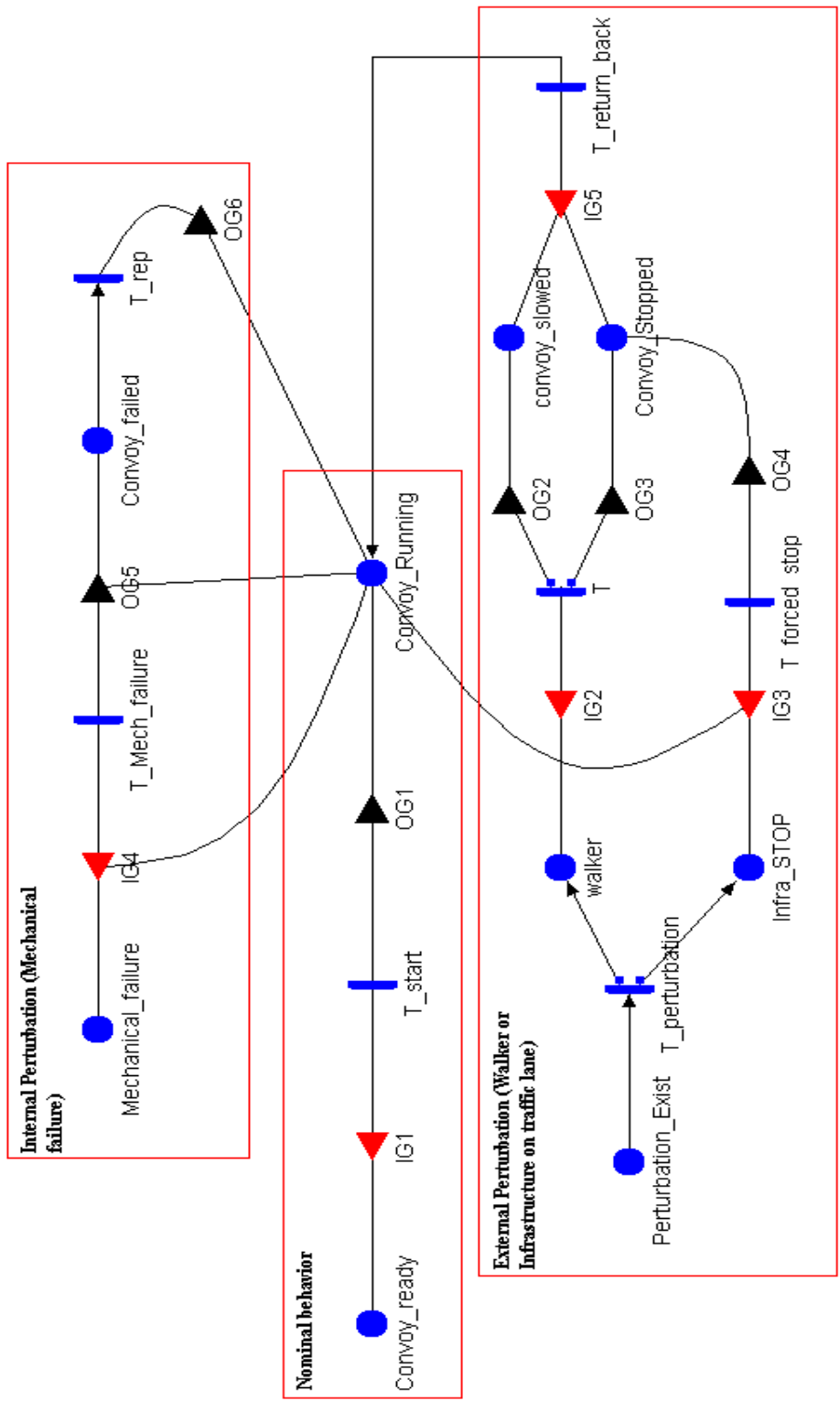

Figure 8. SAN model related to behavior of led convoy in Urban Environment

***Step 6****-Evaluation Measures*

After validating our SAN model for led in convoy, we must determine the measures that we decided to do onto the convoy behavior. In this paper, we will look at the *system availability on the traffic lane in the event of the presence of pedestrians and infrastructure in our environment that is considered dynamic*.

***Step 7****-Quantitative Analysis*

The refined model should now be analyzed by envisaging the chosen evaluation method. This assessment must calculate the metric results. Once the result is obtained, it is necessary to analyze them.

On the quantitative evaluation of our leads in convoy, we chose to use the Möbius tool [8, 9] to perform various measures. We evaluated our system in the presence of a STOP infrastructure in front of the convoy in traffic lane since it is equipped with a dynamic road infrastructure. It is composed of a set of elements that can cause stoppages or slowdowns the vehicles on the road. Lights, intersections can at any time force stop the convoy.

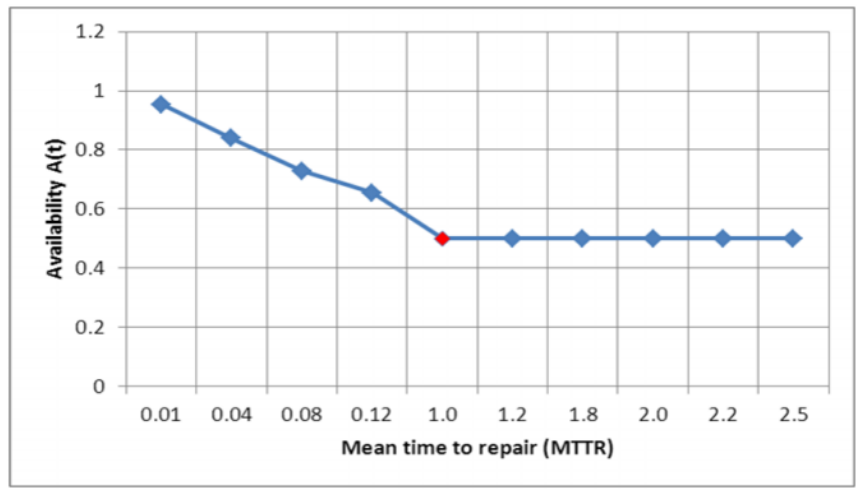

Figure 9.  Effect of MTTR on system Availability

We assessed the system availability A(t) depending the Mean time to repair (MTTR) of the system. We measured the availability of a well-defined time by varying the Mean time to repair of the system.

Fig. 9 shows the system availability evaluation according to the system MTTR in the urban environment. The analysis of results shows that the system Availability decreases in [0.01, 1.0[ and from the MTTR= 1.0, the Availability-Security remains constant value equal to 0.5. This means that it must set the value of the system MTTR to a value greater or equal to 0.1 to obtain the 50% of system Availability.

## 5. CONCLUSIONS

In this paper, we have proposed an Agent-Oriented meta-model adequate to any Transportation Systems with multi-configuration ability problem. The aim is to allow analysts to produce a transportation system model that precisely captures the knowledge and the behavior of an organization so that an agent-oriented requirements specification of the system-to-be.

We illustrated our meta-model on urban public transportation system. We have tried to model our system as multi-agent system based on our proposed meta-model.

Based on proposed meta-model, a methodology for Dependability attributes evaluation Approach of Transportation System in a Dynamic Environment, which consists of several activities. This approach involves the reconnaissance of any convoy, specify the number of each entity involved in the convoy and the clustering strategy and inter-communication between vehicle convoy, set the dependability indices of our system, set the parameters of the convoy that will be examined and used to generate a further analyzable model, generate the formal model describing the behavior of leading in convoy in a specific dynamic environment, chosen dependability indices that must be converted to measures which are calculated from the model and the refined model should now be analyzed by considering the chosen evaluation method (simulation method or analytic method).

Future works will be devoted to several key points aimed at improving the proposed solution. On the one hand, we will work to provide a generic model for a methodology and will be suitable for all platonning application with different scenario in different transportation field. This model is used for the Dependability evaluation. On the second hand, efforts have to be made in order to implement a tool for having an interface to aid the modeling and automate the analysis.

## ACKNOWLEDGEMENTS

Works exposed in this paper are done in collaboration with Systems and Transportation Laboratory (IRTES-SET) with the support of the French ANR (National research agency) through the ANR-VTT Safe platoon project (ANR-10-VPTT-011). University of Technology of Belfort Montbeliard (UTBM), Belfort, France